# Explaining Cost Overruns of Large-Scale Transportation Infrastructure Projects using a Signalling Game

By

Chantal C. Cantarelli, Caspar G. Chorus, and Scott W. Cunningham

## Abstract

Strategic behaviour is one of the main explanations for cost overruns. It can theoretically be supported by agency theory, in which strategic behaviour is the result of asymmetric information between the principal and agent. This paper gives a formal account of this relation by a signalling game. This is a game with incomplete information which considers the way in which parties anticipate upon other parties' behaviour in choosing a course of action. The game shows how cost overruns are the result of an inappropriate signal. This makes it impossible for the principal to distinguish between the types of agents, and hence, allows for strategic behaviour. It is illustrated how cost overruns can be avoided by means of two policy measures, e.g. an accountability structure and benchmarking.

## Introduction

Large-scale projects are often characterised by large cost overruns. Flyvbjerg et al. (2003) conducted an international research of 258 transport infrastructure projects and found that in 86% of the projects under consideration, cost overruns appeared with average cost overruns of 28%. Cost overruns are problematic because they increase the burden on the country's gross domestic product (CBS, 2005 in KiM, 2007). Cost estimates are often inaccurate and consequently the ranking of projects based on project viability is often inaccurate as well. Inevitably, this incorporates the danger that eventually inferior projects are implemented, that resources are used which could have been assigned more appropriately, and that projects are implemented which cannot recover their costs.

Various studies (Hall, 1980; Wachs, 1989; Morris, 1990; Odeck, 2004; Bruzelius et al. 2002; van Wee, 2007; and Flyvbjerg, et al. 2003) addressed this problem of inaccurate cost estimates and provided different accounts for this. First of all, cost overruns are explained by forecasting errors in technical terms, including inadequate data and lack of experience (technical explanations). Secondly, cost overruns are often explained as the result of optimistic forecasts due to cognitive bias (psychological explanations) Next to this, strategic behaviour is an important explanation for cost overruns. It is described as the result of strategic misrepresentation; deliberate underestimation of costs in order to increase the chances for project acceptance (political-economic explanations).

Cost estimates have not improved and overruns have not decreased over the last 70 years (Flyvbjerg et al, 2003). If inaccurate estimates were caused by technical causes, errors in overestimating costs would have been of the same size and frequency as errors in underestimating costs. But this turns out not to be the case. Furthermore, the refinement of data collection and forecasting methods over the years would have resulted in more accurate forecasts over time (Flyvbjerg et al. 2002). However, this was not the case and technical explanations of forecasting errors are, therefore, not considered the main cause for cost overruns. Likewise, because learning effects would have improved the accuracy of the cost estimates if optimism bias is a main reason for underestimation, psychological explanations are not considered the foremost cause of cost overruns. Wachs (1989) found that misleading forecasts of costs were best explained by deception. Flyvbjerg et al. (2002) also concluded that political-economic explanations best fit the data for cost underestimation.

Strategic behaviour seems an important explanation for cost overruns and this will, therefore, be the focus in this paper. There are two main categories of strategic behaviour, adverse selection and moral hazard (Laffont and Tirole, 1998 in Mu et al. 2010). Adverse selection is the tendency that "bad" market parties are selected. It emerges before any contracts are signed in a situation in which the contractor has more information than the owner e.g. regarding the actual costs during a tender process. Moral hazard is strategic behaviour that takes place after the contract is signed. In this case, the contractor takes actions that are not easily observable by the owner such as artificially increasing the project costs during implementation. Both adverse selection and moral hazard are closely related; adverse selection can give rise to moral hazard. The difference between the two concerns who knows what, when. This paper emphasises on the decision-making phase and hence the strategic behaviour of adverse selection, and it considers moral hazard the result of this.

Little work has been done to explain misleading forecasts from a political-economic view. To the authors' knowledge, an explicit application of a theory that illustrates the behaviour of parties leading to cost underestimation has not yet been conducted.

A theory that is particularly suitable to support political-economic explanations is agency theory. Agency theory involves the study where there is a contract in which a client or principal engages an agent or contractor to take actions on behalf of the principal that involve the delegation of some decision-making authority to the agent (Jensen, 2000 in Mu et al. 2010). The theory assumes a relation between the agent and the principal that is characterised by asymmetric information and goal conflict. The agent has more information than the principal and pursues different objectives which might lead to strategic behaviour, the agent not acting in the best interest of the principal. A specific theory that can be used to *formally* describe the behaviour of the principal and agent is game theory (Fudenberg and Tirole, 1992, Rasmussen, 2006). The game that is considered in this paper is a so-called signalling game. It is a game with incomplete information which considers the way in which parties anticipate upon other parties' behaviour in choosing a course of action (Fudenberg and Tirole, 1992).

Game theory's potential to highlight the role of strategic behaviour caused by the asymmetric information between agents makes it a particularly promising framework from the perspective of our research question. This will be subject of this paper.

Previous work has considered the use of multi-attribute analysis techniques for evaluating contractor capability (see for instance Holt et al. 1994). This is useful and complementary work, particularly given the acknowledgment that price is an insufficient signal of contractor quality. The role of price as a signal of quality is further discussed below. Note also that work order or design errors are often a major cause of cost-overruns (Assaf and Al-Hejji, 2006).

However, these play a role mostly after the project has been assigned. In contrast, our study focuses on the process preceding project execution, and specifically on how cost-overruns can be reduced by a more clever design of the tendering process. As such, our study complements –rather than aiming to substitute- studies that deal with cost-overruns from a project-execution perspective. Another related strand of literature focuses on the effect of strategic behaviour during the exploitation of transport infrastructure. For example, Yang and colleagues (2009) use Game Theory to describe the behaviour of private parties, in the absence of a regulative authority, in terms of the exploitation of transport infrastructure by for example setting prices and capacity of toll roads (Yang et al., 2009). In addition, Karlaftis studies how various ownership structures are related to public transit system efficiency (Karlaftis, 2010). This paper, in contrast, focuses on the process that precedes realization of the infrastructure, rather than dealing with the exploitation of the infrastructure.

The paper contributes to the current state of the art on cost overruns of large-scale transportation infrastructure projects as it puts a new perspective on the explanations for cost overruns. The approach of applying game theory herein discerns this study from others. Moreover, game theory can provide a formal account of the interaction between parties, contributing to the scientific underpinning of the explanation. A better theoretical embedded explanation for cost overruns can increase the understanding of strategic misrepresentation of costs by parties and may eventually result in more appropriate measures to deal with this. Two such measurements that can be taken in this respect are considered in this paper. It is shown by means of the signalling game that each of these may influence and improve current practice. The use of game theory therefore not only improves our understanding how cost underestimation occurs but it also shows to what extent measures aimed at dealing with cost-underestimating result in a lower probability of cost overrun as well as in smaller overruns.

Section 2 describes the situation concerning the behaviour of the parties in the process of large-scale transportation infrastructure projects and presents the formal model of the situation. The analysis and implications of the game are discussed in section 4. Section 5 shows how policy measures change the outcome of the game and avoid strategic behaviour by cost underestimation. Section 6 presents the main conclusions and recommendations.

## Specification and Analysis of the Game

A signalling game consists of two players, an informed player (agent) and an uninformed player (principal). The agent has private information that is summarised by his type. The type could for example be his aptitude; his ability to perform the task of the principal. The agent sends to the principal a signal, typically a message that can reveal some of the hidden information of the agent identifying its type. In other words, the message can reveal the extent to which the agent is able to perform the principals' tasks. The principal receives this message and takes an action, after which the game ends. After discussing the game as specified, we demonstrate why project cost is an inadequate measure of contractor capability.

In this paper the following hypothetical situation can be kept in mind when addressing the game. A new road between two cities will be constructed to increase accessibility. It was decided to contract out the realisation of this project. An open tender was organised and market parties submitted their proposal including amongst others the necessary budget. The bid selection criterion is usually based on the lowest costs and this is, therefore, also the selection criterion in this situation. The market party is the agent that will carry out the project according to the wishes of the principal, the governmental authority. The governmental party aims to get the project realised against the lowest costs and the market party aims to get his tender proposal accepted.

Figure 1 depicts the model of the signalling game between the market party and the governmental party.

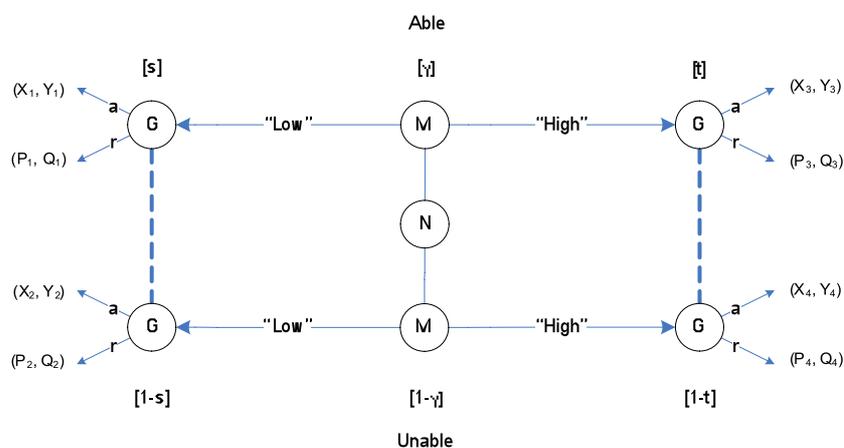

**Figure 1. Signalling game**

The figure is composed of nodes, vector of letters, arrows and labels. Each node is a position in the game; a point at which some player must choose some action. The first position in the game is depicted by an N node; all the other nodes are filled in by either the letter M (Market party) or the letter G (Governmental agency) representing the actor expected to move at that stage of the game. N is the state of nature, which determines a type for the market party, either able or unable. An arrow represents a choice that is feasible for the player choosing.

The game starts with the market party choosing a strategy; either to include a low or high estimate in its tender proposal (Figure 1). The estimate in the tender proposal is the message that the market party sends to the governmental party. The message is the particular choice of the market party between a low or high estimate. The low estimate is the lowest possible realistic estimate perceived as such by the governmental party. The market party's private information is her aptitude: the market party knows whether he is able or unable to realise the project against the low cost estimate. The able market party can realise the project for the low cost estimate whereas the unable market party is not able to realise the project for the low cost estimate. The governmental party will observe the message and has to decide whether to accept (a) or reject (r) the tender proposal of the market party. The arrows a and r point to vectors of letters that are, in this game, composed of two letters (the first letter corresponds to the market party's payoff and the second number refers to the governmental party's payoff). For example, if the able market party provides a low cost estimate in its tender proposal and the governmental party decides to accept the tender, the payoff for the market party is $X_1$ and the payoff for the governmental party is $Y_1$.

The model furthermore indicates the probability that the market party is able by $\gamma$ and indicates the probability that the market party is unable by $1-\gamma$. When making the choice of accepting or rejecting the proposal, the governmental party does not know whether the market party is able or unable (indicated by dashed lines). This is represented in the model as follows. s is the governmental party's belief (probability) that, given that the message with a low estimate is observed, it concerns an able market party. t is the governmental agency's belief (probability) that, given that the message with a high estimate is observed, it concerns an able market party.

**Payoff Structure of the Game**

The payoff structure of the game includes the value of the infrastructure to the governmental party (I). This quantity includes the net benefit of the infrastructure minus minimum cost production of the infrastructure. A second payoff value is the budget requested by the market party for the provision of the infrastructure (either L or H for a low or high estimate respectively). Finally, there is the cost overrun incurred by the unable market party for a low estimate(C). The fall-back options of no infrastructure provision to the government and the market party is fixed to 0.

*Payoff Market Party*

The payoff of the market party is a function of benefits and costs. The benefits include benefits from the budget that is received (either L or H) and additional benefits from receiving commission of the project (R). These additional benefits refer to the reputation of the market party; it is assumed that this reputation increases as the market party becomes more known with each project that is implemented. The magnitude of the additional benefits is independent from the market party's type or message. The costs include costs related to the realisation of the project, which are equal to the cost estimate (thus L or H). In this payoff structure, the payoffs are independent upon the message or type of market party, but the market party prefers acceptance of the proposal above rejection.

*Payoff Governmental Party*

The payoff of the governmental party is a function of benefits and costs. The benefits include the value of the infrastructure (I). These benefits are independent from the market party's type or message. The costs include costs of the budget that is provided to the market party to realise the project (either L of H) and there is cost overrun for the unable market party providing a low estimate.

Figure 2 presents the signalling game including the payoff structure. The payoffs for the market party and governmental party for each strategy set can be determined in this way. For example, if an able market party sends a message with a low cost estimate and the governmental party accepts the tender with this estimate, the payoff for the market party is equal to R and the payoff for the governmental party is equal to I-L.

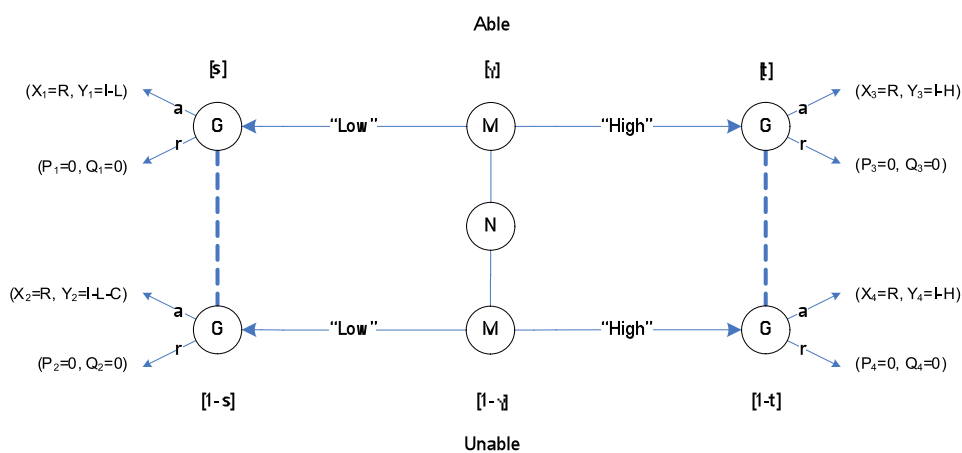

**Figure 2. Signalling game with payoff structure**

**Equilibrium Analysis of the Game**

The strategic form of the game can be useful in deriving the equilibria outcomes of the game taking into account the probabilities of the type of market party (Peters, 2008). Figure 3 gives the computed strategic form of the game.

|  | aa | ar | ra | rr |
|---|---|---|---|---|
| LL | R; $I-L-C$ $(1-\gamma)$ | R; $I-L-C$ $(1-\gamma)$ | $0$ ; $0$ | $0$ ; $0$ |
| LH | R; $I-L(1-\gamma)-H\gamma-C$ $(1-\gamma)$ | $R(1-\gamma)$; $(I-L-C)$ $(1-\gamma)$ | $R\gamma$; $(1-H)\gamma$ | $0$ ; $0$ |
| HL | R; $I-L\gamma-H$ $(1-\gamma)$ | $R\gamma$; $(I-L)\gamma$ | $R(1-\gamma)$; $(1-H)$ $(1-\gamma)$ | $0$ ; $0$ |
| HH | R; $I-H$ | $0$ ; $0$ | R; $I-H$ | $0$ ; $0$ |

**Figure 3. Strategic form of the signalling game**

The market party has the strategy set {LL, LH, HL, HH}, where the first letter refers to the message of an unable market party and the second letter refers to the message of an able market party. L is the strategy of providing a low cost estimate in the proposal and H is the strategy of providing a high estimate in the proposal. The strategy LH means that the unable market party sends a message with a low cost estimate and the able market party sends a message with a high cost estimate.

The governmental party has the strategy set {aa, ar, ra, rr}, where the first letter refers to the action if the market party plays the strategy L and the second letter refers to the action if the market party plays the strategy H. "a" refers to accepting the tender proposal and "r" to rejecting the tender proposal. The strategy "ar" means that if the governmental party receives a message with a low estimate, he will accept the proposal, and if a message with a high estimate is received, he will reject the proposal.

The first figure in the matrix refers to the payoff for the market party and the second to the payoff for the governmental party. E.g. for the strategy set {LL, aa} the payoff for the market party is equal to R, and the payoff for the governmental party is equal to I-L-C (1-γ). In the following material we use the Nash equilibrium concept to solve for potential outcomes or equilibria of the game. Further, we consider whether the government can use the price signal of industry as an indicator of potential contracting quality.

*The Best Strategy of the Sender (Market Party)*

The best response of the sender to the anticipated behaviour of the receiver is the strategy that maximises its utility regardless of the senders' type. In the situation of a tender process, the market party's expected utility is the sum of the probabilities when the action accept/reject is taken when the signal with a low/high estimate is sent multiplied by the utility of the different strategy sets (low/ high, accept/reject, able/unable).

Based on the strategic form of the game presented in Figure 3, the best strategy of the market party can be determined. The best strategy of the market party is determined by considering for each governmental party's strategy the strategy at which the market party receives the highest payoff. For example, considering the anticipated behaviour of the governmental party playing the strategy "aa", the payoffs of the four possible strategies of the market party are compared (LL, LH, HL, HH). In this case, the payoffs for all four strategies of the market party are the same and equal to R.

Considering the anticipated behaviour of the governmental party playing the strategy "ar", the payoffs for LL, LH, HL differ and is highest for the strategy LL (payoff is equal to R whereas the payoff for LH and HL is only a fraction of R, and the payoff for HH is equal to 0). The best strategy of the market party to the anticipated behaviour of the governmental party playing the strategy "ra" is HH (payoff for HH>LH/HL>LL (R>Rγ or R(1-γ)>0)) and for the strategy "rr" each of the four strategies of the market party will provide the same payoff (payoff is 0).

*The Best Strategy of the Receiver (Governmental Party)*

The receiver chooses an action after he observes the sender's message. He wants to make a decision that is optimal given the best beliefs he has concerning the sender's type. The best response of the receiver to the behaviour of the sender is the strategy that maximises its utility. Using the strategic form of the game presented in Figure 3, the best strategy of the governmental party is determined by considering for each market party's strategy the strategy at which the governmental party receives the highest payoff. For example, for the market party's strategy LL the payoffs of the four possible strategies of the governmental party are compared ("aa", "ar", "ra", "rr"). The payoff for the strategies "aa" and "ar" are equal and represented by I-L-C (1-γ) and the payoff for the strategies "ra" and "rr" are equal and represented by 0. Consequently, the best strategy of the governmental party for the market party's strategy LL depends on the probability γ and parameter values I, L and C. With a high probability, γ=1, the best strategy is either "aa" or "ar" if I>L and for a low probability, γ=0, the best strategy is either "aa" or "ar" if I>L+C. Otherwise, the best strategy will be "ra" or "rr". The same applies for the other strategies of the market party, they are all dependent on the probability and parameters values.

Figure 4 shows the strategic form of the game within which the best strategy of the market party and governmental party are indicated.

|  | aa | ar | ra | rr |
|---|---|---|---|---|
| LL | $R^*$; $1-L-C(1-\gamma)$†[1] | $R^*$; $1-L-C(1-\gamma)$†[1] | $0$; $0$† | $0^*$; $0$†[2] |
| LH | $R^*$; $1-L(1-\gamma)-H\gamma-C(1-\gamma)$†[3] | $R(1-\gamma)$; $(1-L-C)(1-\gamma)$† | $R\gamma$; $(1-H)\gamma$† | $0^*$; $0$†[4] |
| HL | $R^*$; $1-L\gamma-H(1-\gamma)$†[5] | $R\gamma$; $(1-L)\gamma$† | $R(1-\gamma)$; $(1-H)(1-\gamma)$† | $0^*$; $0$†[6] |
| HH | $R^*$; $1-H$†[7] | $0$; $0$† | $R^*$; $1-H$†[7] | $0^*$; $0$†[8] |

Note: * indicates the best strategy of the market party, † indicates the best strategy of the governmental party. In this case, all strategies are marked with an † because the best strategy is conditional upon the value of γ and the value of the parameters.
The numbers refer to the outcomes of the game, in which both market party and governmental party play their best strategy

**Figure 4. Strategic form of the signalling game including best strategies**

The computed strategic form of the game presented in Figure 4 identified eight possible equilibria outcomes of the game (these are the situations in which both the market party as the governmental party play their best strategy, indicated by * or † in Figure 4. ). The game will play out differently according to the relative values of I, H and C. L can be recognised as some fraction of H, and therefore only three parameters are addressed rather than 4.

Figure 5 shows a mixture diagram (Png, 1983) indicating which Nash equilibria potentially exist depending on the values of I, H, and C. There are three equilibrium regions with, in each region, four possible equilibria. Regions one and two are associated to variable degrees with cost overruns, whereas region three is associated to variable degrees with a failure to contract infrastructure. This latter is also an interesting problem, but it is not considered here.

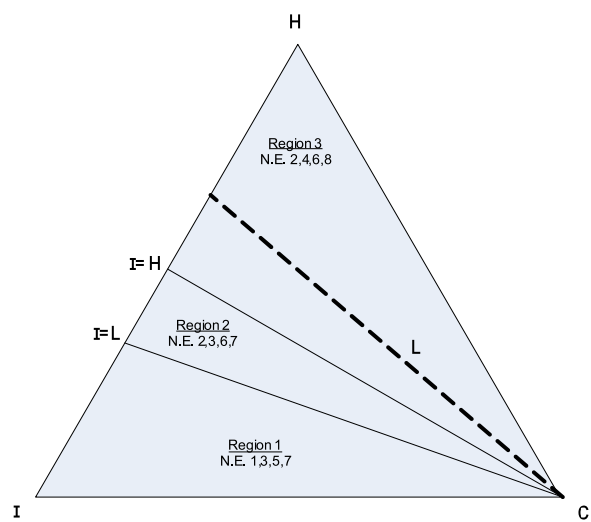

**Figure 5. Mixture diagram identifying 3 regions with in each region different Nash Equilibria (N.E.)**

Market outcomes depend strongly upon the capability of the market. Lower capability markets increase the overall potential for cost overruns. A market with a lower competence may also require additionally money to successfully complete the project. Alternatively, they may attempt to reduce their bid, but be unable to offset the possible additional expected costs of an over-run. A failure to contract or higher overall costs is the resultant outcome. This can be demonstrated analytically by use of extrema – examining how equilibria regions change as competence varies from $\gamma = 0$ to $\gamma = 1$.

## Relation to Cost Overruns

Cost overruns only occur if the governmental party accepts a tender proposal with a low estimate from an unable market party. Two of the equilibrium outcomes identified in the previous section represent this situation, namely, the equilibria 1 and 3. For these equilibria, maximum cost overruns being as high as the government's assessment of the value of the infrastructure (I=C).

### Occurrence of Cost Overruns

Equilibrium 1 {LL, aa} is a so-called *pooling* equilibrium. This is problematic because in this type of equilibrium, both types of market party send the same message and, hence, the message does not reveal anything about the market party's type. The message can be considered an insufficient signal of the game. As a consequence, the governmental party is unable to update his beliefs about the market party's type after receiving a message. The inability of the governmental party to distinguish between the types of market party gives the unable market party the possibility to behave strategically and send a false signal, a signal that does not fit his type, to get the proposal accepted. Thus, despite both parties acting rationally and according to their best response to the other parties' behaviour; whether or not the outcome is desired for society at large depends on the type of market party.

Equilibrium 3 {LH, aa} is a so-called *separating* equilibrium; both types of market party send a different message, which makes it possible to distinguish between the market parties. Despite this, the equilibrium is equally problematic as equilibrium 1. It involves a "bluffing" market party who is in fact unable to complete the contract, coupled with a government willing to accept any bid.

A separating equilibrium can potentially provide useful information to the government in selecting a contractor. Capable contractors may find it advantageous to adopt a specific bidding strategy. The government, upon seeing this strategy used in a given contracting setting, can use the bid as an additional piece of information in comprehensively evaluating contractor capability. A Bayesian approach for handling new information is appropriate in this case. The strategic equivalent for use in a game is known as a perfect Bayesian equilibrium (Ratliff, 1996; Lafont and Tirole, 1998). The use of price as a signifier of competency is credible if and only if there are the correct incentives in the game for a player to reveal their type. In the material which follows we consider whether or not these incentives for open communication actually exist in the game as specified. The necessity of further policy measures is suggested.

**Prevention of Cost Overruns Within the Current Play of the Game**

In the following section we discuss two attempts to remediate cost overruns without making a substantial change to the play of the game. We investigate, and eliminate, the possibility of meaningful pricing signals whereby the competent market party signals their capability thereby enhancing the range of contractible outcomes. We also consider a market covenant where the government promises to pay extra if the parties involved offer a full and complete accounting of cost. This last strategy, known as a "strategic move" in the sense of Schelling (1960), seems promising. However as will be discussed, it is a credible arrangement only for high competence markets undertaking high valued infrastructure.

A price signal can be informative only if the less capable market parties were to play a different strategy than the capable market parties. Further the move must be in equilibrium – compatible with the incentives of both market party and government. However, the least capable market parties will not play a distinct strategy from the capable market players. There is never an incentive for them to do so regardless of government strategy or any degree of prior belief in overall market competence outside g=1 (complete competence). Consider the following reasoning.

It is not possible to induce a meaningful price signal with government strategy "accept all" {aa} since given this response all types of market players play all strategies. Market strategy "reject all" {rr} is doubly problematic; the market signal is informative, and the potential for contracting infrastructure is anyhow lost. The "accept low bid" {ar} strategies induce the market to uniformly bid low and the "accept high bid" strategies {ra} induce the market to uniformly bid high. Here again the bidding response is uninformative. A government strategy to selectively reject low bids is compatible with their interests only when the market is already understood to be completely capable. The signal is both uninformative and uninteresting. A similar argument eliminates the possibility of selectively accepting high bids.

Another option would be for the market parties to communicate outside the game. We discuss below a covenant which might be reached between high competency markets and the government. Suppose the government promised to unconditionally accept high bids {aa}, as long as the least capable parties makes a full and complete accounting of their costs {HL}. Such a prior commitment would be credible to both parties only if (a) the government could credibly claim that it could cost less money for it to accept some high bids than to rejecting all high bids, and (b) the contracted infrastructure was worth the extra cost to the government.

Requirement a is trivially satisfied for all cases where the government might want to contract, I > L. Requirement a however puts some binding constraints on the minimum necessary level of market competency. Such a claim from the government is credible if the strategy of {aa} is more valuable to the government than the strategy {ar}, subject to industry promising to play {HL}. The claim results in an inequality, which working through can be expressed as a constraint on the required competence of the market.

$$\gamma > ((H-I))/((H-L)).$$

The greater the high cost bid relative to infrastructure, the greater the need for market competency for this covenant to be credibly upheld. Likewise, the more efficient the low bid, relative to infrastructure, also the greater the requirement on market competency. The covenant, while attractive, is possible only to more developed and capable marketplaces.

## The Influence of Policy Measures

The previous sections described by means of a signalling game how cost overruns could occur. It showed that market parties were able to behave strategically by underestimating costs due to the lack of an appropriate signal to distinguish between the market party's types. In order to prevent market parties underestimating costs, the incentive structure has to change in such a way that the signal becomes effective. This will be further elaborated upon in this section. Two policy measures will be addressed in this respect, the introduction of an accountability structure and the introduction of a benchmark system.

### Introducing an Accountability Structure

The accountability structure refers to the way in which risks are distributed between parties. The governmental party is usually responsible for most of the risks in projects, including cost overruns. With a different accountability structure (for example in the form of a Public-Private Partnership or alliance contracting setting) part of the risk for cost overruns can be transferred to the market party.

Considering the market party's payoff, being a function of benefits and costs. The accountability for cost escalations is included in the game by means of a modification of the payoff structure, i.e. next to the costs related to the realisation of the project, additional costs are made in case the market party is not able to realise the project for the provided budget (C). The market party is responsible for a fraction f of the total cost overruns, and consequently, the governmental party is responsible for fraction 1-f of the total cost overruns. This is the case if the unable market party provides a low estimate.

The best strategy of the market party is determined by the strategy with the maximum utility over all other strategies. Since this measurement only affects the unable market party, the best strategy of the able market party will remain the same. The payoffs for the unable market party are reduced by the additional costs in case of cost overruns.

Figure 6 presents the strategic form of the game with the accountability structure.

|  | aa | ar | ra | rr |
|---|---|---|---|---|
| LL | $R\text{-}Cf(1\text{-}\gamma)$ ; $1\text{-}I\text{-}C(1\text{-}\gamma)(1\text{-}f)^c$ | $R\text{-}Cf(1\text{-}\gamma)$ ; $1\text{-}I\text{-}C(1\text{-}\gamma)(1\text{-}f)^c$ | $0$; $0f$ | $0^*$; $0f^{11}$ |
| LH | $R\text{-}Cf(1\text{-}\gamma)$ ; $1\text{-}I\text{-}(1\text{-}\gamma)\text{-}H\gamma\text{-}C(1\text{-}\gamma)(1\text{-}f)^c$ | $(R\text{-}Cf)(1\text{-}\gamma)$; $(1\text{-}I\text{-}C(1\text{-}f))(1\text{-}\gamma)^c$ | $R\gamma$; $(1\text{-}H)\gamma^c$ | $0^*$; $0f^{12}$ |
| HL | $R^*$; $1\text{-}I\text{-}\gamma\text{-}H(1\text{-}\gamma)^{f3}$ | $R\gamma^*$; $(1\text{-}I)\gamma^{f4}$ | $R(1\text{-}\gamma)$; $(1\text{-}H)(1\text{-}\gamma)^c$ | $0^*$; $0f^{15}$ |
| HH | $R^*$; $1\text{-}H^{f6}$ | $0$; $0f$ | $R^*$; $1\text{-}H^{f6}$ | $0^*$; $0f^{17}$ |

**Figure 6. Strategic form of the game with an accountability structure (general)**

In the starting situation, considering the anticipated behaviour "aa" by the governmental party, all four strategies resulted in equal expected payoffs for the market party. The introduction of an accountability structure will affect the payoff for the unable market party if he provides a low estimate and hence, only the expected payoff for the strategies LL and LH will change. The expected payoff for these strategies will decrease ($X_2 < X_{1=3=4}$) and the best strategy for the market party is, therefore, HL or HH (the payoffs for these strategies are equal since $X_{1=3=4}$). Similarly, the best strategies of the market party for the other behaviours of the governmental party ("ar", "ra" and "rr") are determined and marked with an asterisk in Figure 6. The best strategies of the market party will not change for the behaviour strategies "ra" or "rr" of the governmental party since they both reject the low estimate and the accountability structure is not part of the game. The best strategy of the governmental party will remain dependent on the values of I, C and H and the fraction f.

The game has 7 equilibrium outcomes. Equilibria 1, 2, 5 and 7 concern the failure to contract infrastructure and will not be considered here.

- N.E. 3 {HL, aa}: the market party sends the message that fits it type and the governmental party accepts all bids. This involves inefficiency because the governmental party is still involved in contracting with unable market parties.
- N.E. 4. {HL, ar}: the market party sends the message that fits its type and the governmental party accept only the low bid from the able market party and a wastage of financial resources (by accepting a high bid from an unable market party) is avoided.
- N.E. 6 {HH, aa} and {HH, ra}: both market parties provide high estimates and the governmental party accepts the proposal. This could lead to financial wastage as market parties could provide unrealistic high bids anticipating upon the governmental party's strategy of accepting the high bid

The accountability structure eliminates both problematic equilibria associated with cost overruns. A new separating equilibrium {HL, ar} emerges, in which the market party sends the message that fits its type and the governmental party only accepts the low estimate by the able market party. The accountability structure prevents the unable market party from providing a low estimate and the governmental party can accept a proposal with a low estimate without the danger of cost overruns.

**Introducing a Benchmark System**

Benchmark systems compare the performance of companies and integrate this information into the selection of firms for future contracts. The introduction of a benchmark system provides the governmental party with information on the past performance of market parties in the tender process. The additional information that is made available to the governmental party reduces the information asymmetry between the governmental party and the market party and limits in this way the possibilities of strategic behaviour by market parties.

With a benchmark system, the governmental party faces, in addition to the variability uncertainty regarding the market party's type, uncertainty regarding the quality of the benchmark. This can be represented in the game by a second signal to the governmental party concerning the quality of the market. This signal tells whether the estimate reveals the truth about the market party's type. Figure 7 presents this game. The probabilities of the benchmark system are indicated by q and r for the able and unable market party respectively. The benchmark system changes the expected utility of the different strategies due to the conditional probabilities of the decision nodes representing the benchmark signal.

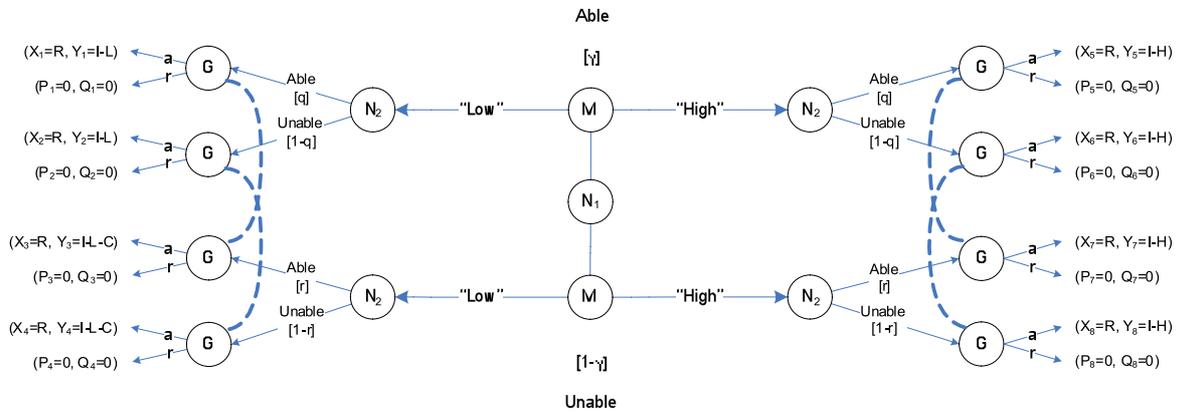

**Figure 7. Signalling game for the benchmark system including two signals**

Figure 8 presents the strategic form of the game. The additional signal increases the number of possible strategies for the governmental party. The strategy set can be summarised by 4 letters, the first letter refers to the governmental party's action (accept or reject) if he receives a low estimate and the benchmark identifies the market party as unable, the second refers to his action if he receives a low estimate and the benchmark identifies the market party as able, the third and fourth letter refer to the actions when he receives a high estimate and the benchmark system identifies the market party as unable and able respectively. Note that there are a number of strategic equivalent equilibria in the payoff matrix, e.g. the strategies {aara}, {aaar}, and {aarr} are strategically equivalent equilibria to the strategy {aaaa} and are not further mentioned.

| | LL | LH | HL | HH |
|---|---|---|---|---|
| aaaa | $R$ (*) <br> $I-L-C(1-\gamma)$ (†) | $R$ (*) <br> $I-L-(1-\gamma)-C(1-\gamma)-H\gamma$ (†) | $R$ (*) <br> $I-H(1-\gamma)-L\gamma$ (†) | $I-H$ <br> (†) |
| raaa | $R(\gamma q+(1-\gamma)r)$ <br> $(I-L)\gamma q+(I-L-C)(1-\gamma)r$ | $R(\gamma+(1-\gamma)r)$ <br> $(I-L-C)(1-\gamma)r+(I-H)\gamma$ | $R(\gamma q+(1-\gamma))$ <br> $(I-L)\gamma q+(I-H)(1-\gamma)$ | $R$ <br> $I-H$ (†) |
| araa | $R((1-r)+ \gamma(1-q))$ <br> $(I-L)\gamma(1-q)+(I-L-C)(1-\gamma)(1-r)$ | $R(1-r+r\gamma)$ <br> $(I-L-C)(1-\gamma)(1-r)+(I-H) \gamma$ | $R(1-\gamma q)$ <br> $(I-L)\gamma(1-q)+(I-H)(1-\gamma)$ | $R$ <br> $I-H$ (†) |
| aara | $R$ (*) <br> $I-L-C(1-\gamma)$ (†) | $R(\gamma q+(1-\gamma))$ <br> $(I-L-C)(1-\gamma)+(I-H)\gamma q$ | $R(\gamma q+(1-\gamma)r)$ <br> $(I-L)\gamma+(I-H)(1-\gamma)r$ | $R(\gamma q+(1-\gamma)r)$ <br> $(I-H)( \gamma q+(1-\gamma)r)$ |
| aaar | $R$ (*) <br> $I-L-C(1-\gamma)$ (†) | $R(1-\gamma q)$ <br> $(I-L-C)(1-\gamma)+(I-H)\gamma(1-q)$ | $R(1-r+r\gamma)$ <br> $(I-L)\gamma+(I-H)(1-\gamma)(1-r)$ | $R((1-\gamma)(1-r)+ \gamma(1-q))$ <br> $(I-H)(\gamma(1-q)+(1-\gamma)(1-r))$ |
| rraa | $0$ <br> $0$ | $R\gamma$ <br> $(I-H) \gamma$ | $R(1-\gamma)$ <br> $(I-H)(1-\gamma)$ | $R$ <br> $I-H$ (†) |
| rara | $R(\gamma q+(1-\gamma)r)$ (*) <br> $(I-L)\gamma q+(I-L-C)(1-\gamma)r$ (†) | $R(\gamma q+(1-\gamma)r)$ (*) <br> $(I-L)(1-\gamma)r+(I-H)\gamma q$ (†) | $R(\gamma q+(1-\gamma)r)$ (*) <br> $(I-L)\gamma q+(I-H)(1-\gamma)r$ (†) | $R(\gamma q+(1-\gamma)r)$ (*) <br> $(I-H)(\gamma q+(1-\gamma)r)$ (†) |
| raar | $R(\gamma q+(1-\gamma)r)$ <br> $(I-L)\gamma q+(I-L-C)(1-\gamma)r$ | $R((1-r)+(1-q)\gamma)$ <br> $(I-L-C)(1-\gamma)r+(I-H)\gamma(1-q)$ | $R((\gamma q+(1-\gamma)(1-r))$ (*) <br> $(I-L)\gamma q+(I-H)(1-\gamma)(1-r)$ (†) | $R((1-\gamma)(1-r)+ \gamma(1-q))$ <br> $(I-H)(\gamma(1-q)+(1-\gamma)(1-r))$ |
| arra | $R((1-\gamma)(1-r)+\gamma(1-q))$ <br> $(I-L)\gamma(1-q)+(I-L-C)(1-\gamma)(1-r)$ | $R(\gamma q+(1-\gamma)(1-r))$ (*) <br> $(I-L-C)(1-\gamma)(1-r)+(I-H)\gamma q$ (†) | $R(\gamma(1-q)+(1-\gamma)r)$ <br> $(I-L)\gamma(1-q)+(I-H)(1-\gamma)r$ | $R(\gamma q+(1-\gamma)r)$ <br> $(I-H)(\gamma q+(1-\gamma)r)$ |
| arar | $R((1-\gamma)(1-r)+ \gamma(1-q))$ <br> $(I-L)\gamma(1-q)+(I-L-C)(1-\gamma)(1-r)$ | $R((1-\gamma)(1-r)+ \gamma(1-q))$ (*) <br> $(I-L-C)(1-\gamma)(1-r)+(I-H)\gamma(1-q)$ (†) | $R((1-\gamma)(1-r)+ \gamma(1-q))$ (*) <br> $(I-L)\gamma(1-q)+(I-H)(1-\gamma)(1-r)$ (†) | $R((1-\gamma)(1-r)+ \gamma(1-q))$ <br> $(I-H)(\gamma(1-q)+(1-\gamma)(1-r))$ |
| aarr | $R$ (*) <br> $I-L-C(1-\gamma)$ (†) | $R(1-\gamma)$ <br> $(I-L-C)(1-\gamma)$ | $R\gamma$ <br> $(I-L) \gamma$ | $0$ <br> $0$ |
| arrr | $R((1-\gamma)(1-r)+ \gamma(1-q))$ (*) <br> $(I-L)\gamma(1-q)+(I-L-C)(1-\gamma)(1-r)$ (†) | $R(1-\gamma)(1-r)$ <br> $(I-L-C)(1-\gamma)(1-r)$ | $R\gamma(1-q)$ <br> $(I-L)\gamma(1-q)$ | $0$ <br> $0$ |
| rarr | $R(\gamma q+(1-\gamma)r)$ (*) <br> $(I-L)\gamma q+(I-L-C)(1-\gamma)r$ (†) | $R(1-\gamma)r$ <br> $(I-L-C)(1-\gamma)r$ | $R\gamma q$ <br> $(I-L)\gamma q$ | $0$ <br> $0$ |
| rrar | $0$ <br> $0$ | $R\gamma(1-q)$ <br> $(I-H)\gamma(1-q)$ | $R(1-\gamma)(1-r)$ <br> $(I-H)(1-\gamma)(1-r)$ | $R((1-\gamma)(1-r)+ \gamma(1-q))$ (*) <br> $(I-H)(\gamma(1-q)+(1-\gamma)(1-r))$ (†) |
| rrra | $0$ <br> $0$ | $R\gamma q$ <br> $(I-H)\gamma q$ | $R(1-\gamma)r$ <br> $(I-H)(1-\gamma)r$ | $R(\gamma q+(1-\gamma)r)$ (*) <br> $(I-H)(\gamma q+(1-\gamma)r)$ (†) |
| rrrr | $0$ (*) <br> $0$ (†) | $0$ (*) <br> $0$ (†) | $0$ (*) <br> $0$ (†) | $0$ (*) <br> $0$ (†) |

Note: for reasons of readability, the columns represent the strategies of the market party and the rows represent the governmental party's strategies. Furthermore, the first row in the cell represents the expected payoffs for the market party and the second row in the cell represents the expected payoff for the governmental party; e.g. the strategy set {LL, aa}, R is the expected payoff for the market party, and I-L-C(1-γ) is the expected payoff for the governmental party.

**Figure 8. Strategic form of the game with a benchmark system**

The benchmark system includes two type or errors. The first type of error concerns the benchmark system telling that the market party is able but it is in fact unable. The probability of this type of error is represented by the probability r. The second type of error concerns the benchmark system telling the market party is unable but it is in fact able. The probability of this type of error is represented by 1-q. Type 1 errors are much worse as compared to type 2 errors and the probability of occurrence should, therefore, be reduced. At any stake, the benchmark system should be such that 1-q<0.5 and r<0.5. Otherwise, the benchmark would be misleading and not helpful. Based on these constraints, the best strategy of the market party can be determined (marked with an asterisk in Figure 8).

To determine the best strategy of the governmental party, phased diagrams are used. Figure 9 provides an example of such phase diagrams. The two most interesting strategies of the market party, the strategies in which cost overruns could occur, are illustrated. The dashed line represents the best strategy of the governmental party for the strategy LL or LH by the market party for different market competence. The shaded regions indicate the benefits as a result of the benchmark system.

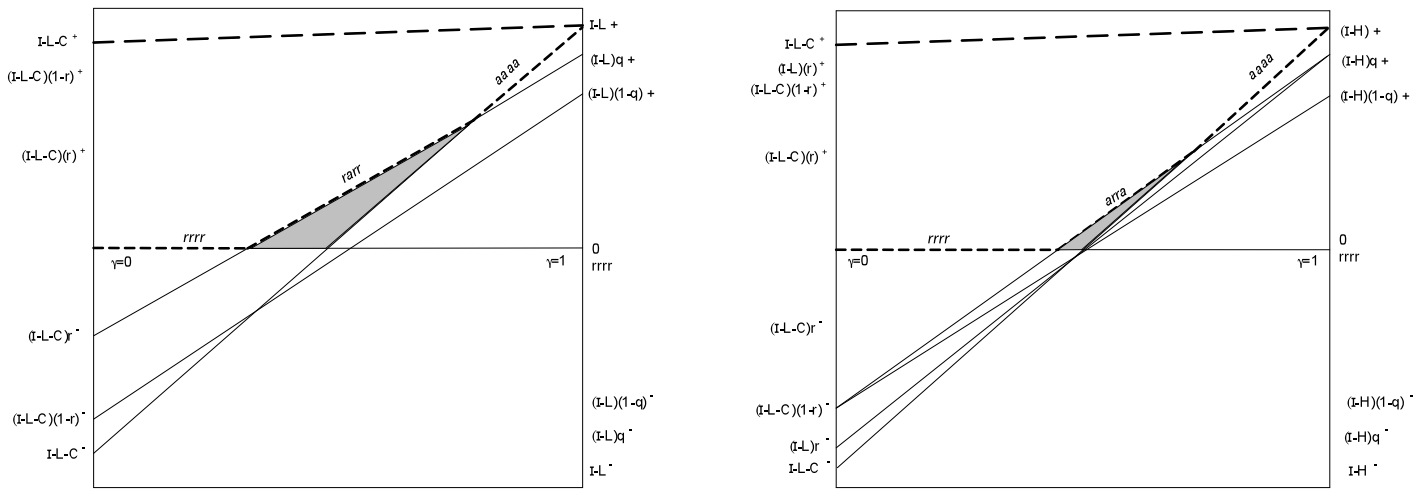

**Figure 9. Phase diagram for market party's strategy {LL} (left diagram) and {LH} (right diagram)**

The analyses by phase diagrams demonstrate the following.

Case 1: no deficits associated with overrun or high bids:

- {aaaa} dominates for all strategies of the market party

Case 2: deficits associated with overruns but not with high bids:

- {LL}{rrrr} or {LL}{rarr} or {LL}{aaaa} dominates depending upon market competence
- {LH}{rrrr} or {LH}{arra} or {LH}{aaaa} dominates depending upon market competence
- {HL}{aaaa} dominates
- {HH}{aaaa} dominates

The benchmarking provides additional strategic flexibility to manage a more complete range of market competencies. Benchmarking is of very considerable help when the market is bidding low {LL} since it enables selection of competent market parties. The benchmark introduces two additional best strategies for the governmental party {rara} and {rarr} next to the best strategies in the starting situation {aaaa} and {aarr}. The new enabled strategy {rara} rejects the low as well as high estimate if the benchmark tells it concerns an unable market party. It is, therefore, a market disciplinary strategy for high bidders as well but it does not apply in the game of cost overruns. In this strategy, the governmental party accepts high bids from the able market party, and this would result in financial wastage because the market party is able to realise the project for less. Therefore, although both new enabled strategies are equivalent, the strategy {rarr} is preferred; only bids from an able market party should be accepted.

In the new enabled strategy {rarr}, the market party bids low and the government uses benchmarking to accept only bids from market parties with a certain level of market competence. Thus, the benchmarking enables more effective contracting in lower capability markets (shaded area in Figure 9).

A benchmark system is of only limited use when the market parties are bluffing {LH}. The benchmarking introduces one additional best strategy for the governmental party i.e. {arra}. This strategy affords a credible threat for the government to penalise and possibly eliminate a subset of moderately competent market parties. These parties are subjected to an error-laden benchmarking system, while the most competent are fast-tracked with an immediate and unconditioned acceptance.

**Comparison Between the Accountability and Benchmark System**

Both the accountability as well as the benchmark system reduce the probability of strategic behaviour by the market party. This section further elaborates upon the advantages and disadvantages of both systems. The accountability system removes the incentive to provide underestimated costs (the unable market party providing a low estimate) because the market party is held responsible for any additional costs and his behaviour is reprimanded if he cannot realise the budget against the agreed budget. The change in the payoff structure of the game changes the behaviour of the market party. The government has a large advantage of this system because he is not being confronted with any large unforeseen cost overruns.

There are, however, two possible dangers with the introduction of an accountability structure. First of all, although it reduces the strategic behaviour in the decision-making phase, it can give rise to another type of strategic behaviour during the implementation phase, so-called moral hazard. With moral hazard, the danger exists that the market party is trying to stay within the budget in order to avoid any additional costs that he is now responsible of, at the expense of other project values. The information asymmetry between the market party and the governmental party makes it possible for the market party to reduce the performance on these other project values e.g. scope or quality because the governmental party is not able to fully monitor the market party. Secondly, care should be taken into determining the extent to which responsibilities are transferred to the market party. Too large risks and responsibilities for the market party will result in yet another type of strategic behaviour that is contradictory to the current situation, i.e. market parties will provide unrealistic high bids to ensure that they can realise the project for the received budget.

The benchmark system does not, contrary to the accountability structure, alter the payoff structure to reduce strategic behaviour. It does so by reducing the information asymmetry between parties by providing additional information to the governmental party. This makes the governmental party's belief about the market party's type more certain which decreases the probability of an inappropriate action with unforeseen cost overruns. The effectiveness of the benchmark system is, however, dependent upon the market party's strategy. It is not fully effective for a "bluffing" market party, thus still incorporating the danger of potential foreseen cost overruns. The government, similar as with the accountability system, has an advantage of the system because he is not confronted with any unexpected cost overruns. Next to this, the increased transparency of the system enables him to make a decision based on complete information. However, the benchmark system also has its demerits. Although it is introduced to reduce strategic behaviour of the market party underestimating costs, it can at the same time evoke strategic behaviour of a different kind. It concerns the danger of so-called *signal jamming*. Signal jamming concerns the tendency of the market party to hide information for the governmental party as to not to reveal their true type (Rasmussen, 2006).

## Conclusions and Recommendations

### Main Conclusions

This paper focuses on political-economic explanations for cost overruns. These explain cost overruns as the strategic misrepresentation of costs by agents to get their proposal accepted. The main objective of this research is to illustrate this strategic behaviour from a theoretical perspective. The theory of principal and agent is often used in this respect. This paper uses game theory, and more specifically a signalling game to address the principal-agent problem by providing a formal account of the interaction between parties that result in the underestimation of costs. It shows how the market party and the governmental party choose a course of action that maximises their expected payoff through anticipation of behaviour.

The paper shows that there are multiple equilibria outcomes in a simple model of cost overruns. The problematic equilibria concern the situation in which the governmental party accepts the low estimate from the unable market party. This situation can emerge due to a lack of an appropriate incentive for the market party to provide the message that fits its type. Furthermore, this strategic behaviour is possible because the signal in the game is insufficient for the governmental party to distinguish, based on the message he receives, between the type of market party and to accept only a low estimate from the able market party.

The signalling game gives useful insights in the way in which strategic behaviour results in cost underestimation. It is, furthermore, a valuable tool to predict the impact of policy measures on the behaviour of the market party. Measurements are aimed to reprimand or prevent the strategic behaviour of the market party and they should be focused on changing the incentive structure in such a way that the signal of the game becomes effective. Two such measurements are considered in this paper, i.e. the introduction of an accountability structure and a benchmark system. Overall, it was shown that the measurements have the desired impact of reducing the probability of cost overruns, but they can also give rise to other kinds of strategic behaviour such as moral hazard or cost overestimation. These problems can also be modelled by means of a signalling game in the same way as cost underestimation was modelled in this paper.

**Limitations of Our Model**

As with many studies of signalling games, this study suffers from certain limitations. This is the result of assumptions and simplifications necessary to capture the situation in the game. This section addresses these limitations by considering for each assumption and simplification the way in which it affects the game and outcome.

This paper uses the tender price as the bid selection criterion (represented by a message that is based on costs). Strategic behaviour occurs due to asymmetric information about the selection criteria. However, the market party only has strategic advantage over the governmental party concerning the actual costs. The information asymmetry with respect to the other selection criteria is small and consequently, opportunities for strategic behaviour regarding these criteria are fewer. The outcome of the game is, therefore, mainly determined by the extent to which the market party behaves strategically regarding the cost estimate rather than to other criteria. A signalling model with one message regarding the estimated costs, therefore, suffices. The assumption to focus on one bid selection criterion does not set any restrictions to the conclusions of this research.

The message in the game is binary but the estimated cost is actually a continuous variable. Modelling the variable accordingly will result in the same outcome and shows, in addition, the extent of underestimation. However, this paper addresses the tendency towards cost underestimation and not the extent to which costs are underestimated.

The market party is either one of the following two types, an able or unable market party. Market parties differ to the extent to which they are able to realise the project for the low cost estimate. Differences in the extent of ability will influence the extent of cost overruns but this is not considered in this research. Concerning the payoff function it is assumed that the payoff is a linear function of the costs and benefits. The assumption of linearity influences this ratio between the payoffs and equilibrium outcome. It mainly affects the tipping points at which one strategy is still better than another..

**Avenues for Further Research**

The formal model presented in this paper presents three benefits for continued research. First, it suggests areas for empirical investigation and testing. Second, it provides a theoretical explanation which assists the choice and justification of specific remedies for alleviating overruns. Third, it contributes to a growing body of evidence that overruns occur, at least in part, as a result of strategic behaviour. We review each of these points in greater detail below.

The model provides firm hypotheses which can be empirically verified. It calls specific attention to the role of market capabilities, the spread between high and low bids, and the value of the infrastructure as potential determinants of cost overruns. The model also demonstrates that there is an important source of censoring in empirical data – the failure to contract desired infrastructure. A complete database should include those projects which are completed, as well as those which fail in contracting.

The model presents a clear and falsifiable hypothesis concerning the maximum degree of cost over-run. In the model, overruns cannot exceed 100% of the valued infrastructure. This is falsified in at least one notable case (Murphy 2008). Cases such as these suggest interesting extensions to the model; the ultimate answer seems to require more strategic behaviour rather than less. The model also presents a clear and falsifiable hypothesis concerning market competency. Cost overruns should occur within regimes corresponding to local markets for capability and competency. Two counter claims could be tested; that cost overruns are not regionally specific, or that cost overruns are regionally specific but occur solely from a lack of engineering competency rather than strategic misrepresentation.

Flyvbjerg et al. (2003) have provided a number of instruments to strengthen accountability. These measures have been applied in practice in policy and planning to address misinformation or strategic behaviour of underestimating costs, and signs of improvement have recently appeared. Although these measures are focused on reducing misinformation by project promoters instead of by market parties, the rationale is the same, strategic behaviour by underestimating costs to get the project approved or tender accepted respectively. We therefore believe that the effects will also be considerable for the situation drawn in this paper. Whether this is actually the case can be tested by comparing the cost performance of projects with different levels of accountability.

The measurement of benchmarking has also been applied in practice to address strategic misrepresentation or underestimating of costs, by the method of reference class forecasting (the theory was developed by Kahneman and Tversky and the method and data were developed by Flyvbjerg). "Reference class forecasting consists of taking a so-called "outside view" on the particular project being forecast. The outside view is established on the basis of information from a class of similar projects" (Flyvbjerg, 2009 – survival of the unfittest). The method has proven more accurate than conventional forecasting, and has been applied by governments and private companies over the world. Similar to the measurement of accountability, the method is focussed on strategic behaviour of project promoters but it can just as well be applied for the strategic behaviour of market parties. For this purpose, data should be gathered about the past performance of market parties in tenders. This information can be used as a reference to establish the accurateness of the estimate or the ability of the market party.

The covenant approach discussed in the paper is also currently being applied in practice. One recent example involves the Dutch *Ministry of Infrastructure and Environment* which has adopted a new memorandum of understanding regarding high bids in infrastructure projects. The model presented in this paper makes it clear that given the game structure considered such covenants can only work under certain specific situations involving high competency and manageable infrastructure costs.

## Acknowledgements

The research was supported by the Dutch Ministry of Infrastructure and Environment. We are grateful to three referees, whose thoughtful comments have been instrumental in improving an earlier version of this paper.

## List of Figures